\documentclass[11pt]{article}
\usepackage{moriond,epsfig}

\newcommand{\bdk}{$B^{\pm}\to DK^{\pm}$}
\newcommand{\bdtk}{$B^{\pm}\to \tilde{D}K^{\pm}$}

\newcommand{\bdsk}{$B^{\pm}\to D^{*}K^{\pm}$}
\newcommand{\bdstk}{$B^{\pm}\to \tilde{D}^{*}K^{\pm}$}

\newcommand{\bddsk}{$B^{\pm}\to D^{(*)}K^{\pm}$}
\newcommand{\bddstk}{$B^{\pm}\to \tilde{D}^{(*)}K^{\pm}$}

\newcommand{\bdpi}{$B^{\pm}\to D\pi^{\pm}$}
\newcommand{\bdtpi}{$B^{\pm}\to \tilde{D}\pi^{\pm}$}

\newcommand{\bdstpi}{$B^{\pm}\to \tilde{D}^{*}\pi^{\pm}$}

\newcommand{\bddstpi}{$B^{\pm}\to \tilde{D}^{(*)}\pi^{\pm}$}

\newcommand{\dsdpi}{$D^{*\pm}\to D\pi_s^{\pm}$}

\newcommand{\dkpp}{$\bar{D^0}\to K_s\pi^+\pi^-$}
\newcommand{\dtkpp}{$\tilde{D}\to K_s\pi^+\pi^-$}

\begin{document}
\vspace*{4cm}
\title{MEASUREMENT OF \boldmath{$\phi_3$} IN \bddsk\ DECAYS AT BELLE}
\author{A. POLUEKTOV~\footnote{Representing the Belle Collaboration.}}
\address{The Budker Institute of Nuclear Physics\\
Acad. Lavrentiev prospect 11, 630090 Novosibirsk, Russia\\
}
\maketitle
\begin{figure}[h]
\begin{center}
\includegraphics[height=3.5cm]{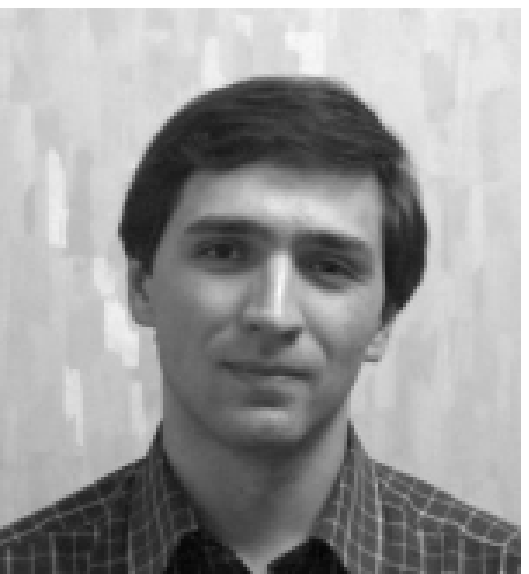}
\end{center}
\end{figure}
\abstracts{
We present a measurement of the unitarity triangle angle $\phi_3$  
using a Dalitz plot analysis of the three-body decay of the $D^0$ meson from 
the \bddsk\ process. 
The method employs the interference 
between $D^0$ and $\bar{D^0}$ to extract both the weak and strong phases. 
We apply this method to the 140 fb$^{-1}$ of data
collected by Belle experiment. 
The analysis uses \bdk\ and \bdsk\ with $D^{*}\to D\pi^0$ modes, where
the neutral $D$ meson decays into $K_s\pi^+\pi^-$. 
From a combined maximum likelihood fit of \bdk\ and \bdsk\ modes, 
we obtain $\phi_3=81^{\circ}\pm 19^{\circ}\pm 13^{\circ}
\mbox{(syst)}\pm 11^{\circ}(\mbox{model})$. The 95\% confidence interval 
is $35^{\circ}<\phi_3<127^{\circ}$. 
}

\section{Introduction}

The determination of the Cabibbo-Kobayashi-Maskawa (CKM) matrix 
elements~\cite{ckm} is important to check the consistency of the Standard Model 
and search for new physics. 
Various methods using $B\to D K$ decays have been introduced~\cite{bdk_gamma}
to measure the unitarity triangle angle $\phi_3$
but the statistics accumulated by currently running experiments is 
not yet sufficient to constrain the value of $\phi_3$ with a reasonable 
significance. 
A novel technique based on the analysis of the three-body 
decay of the $D^0$ meson~\cite{dalitz_bdk} 
has a higher statistical precision compared to the 
methods based on branching fraction measurements.

This method is based on two key observations:  neutral $D^{0}$ and $\bar{D}^{0}$
mesons can decay to a common final state such as $K_s \pi^+ \pi^-$, 
and the decay
$B^+\to D^{(*)} K^+$ can produce neutral $D$ mesons of both flavors 
via $\bar{b}\to \bar{c}u\bar{s}$ and $\bar{b}\to \bar{u}c\bar{s}$ 
transitions, where the relative phase $\theta_+$ between the two interfering
amplitudes is the sum, $\delta + \phi_3$, of strong and weak interaction
phases.  In the charge conjugate mode, the relative phase
$\theta_-=\delta-\phi_3$, so both phases can be extracted 
from the measurements of such $B$ decays and their charge conjugate modes. 
The phase measurement is based
on the analysis of Dalitz distribution of the three body final state of the 
$D^{0}$ meson.

In the Wolfenstein parameterization of the CKM matrix, 
the amplitudes of the diagrams
contributing to the decay $B^+\to \bar{D^0}(D^0)K^+$
are given by dominant $M_1\sim V_{cb}^*V_{us}\sim A\lambda^3$ and
color suppressed $M_2\sim V_{ub}^*V_{cs}\sim A\lambda^3(\rho+i\eta)$
matrix elements.
The two amplitudes interfere if the $D^0$ and $\bar{D}^0$ mesons decay
into the same final state $K_s \pi^+ \pi^-$; 
we denote the admixed state as $\tilde{D}$. Assuming no CP 
asymmetry in $D$ decays, the amplitude of the $B^+$ decay 
is written as
\begin{equation}
\label{intdist}
M_+=f(m^2_+, m^2_-)+re^{i\phi_3+i\delta}f(m^2_-, m^2_+), 
\end{equation}
where $m_{+}^2$ and $m_{-}^2$ are the squared 
invariant masses of the $K_s \pi^+$ and
$K_s \pi^-$ combinations, respectively, and $f(m_+, m_-)$ 
is the complex amplitude of the decay
\dkpp. The absolute value $r$ of the ratio between 
the two interfering amplitudes is given by the product of the ratio 
$|V_{ub}^*V_{cs}|/|V_{cb}^*V_{us}|\sim 0.38$ and
the color suppression factor. 

Similarly, the amplitude of the charge conjugate $B^-$ decay is 
\begin{equation}
\label{intdist_m}
M_-=f(m^2_-, m^2_+)+re^{-i\phi_3+i\delta}f(m^2_+, m^2_-). 
\end{equation}
Once the functional form of $f$ is fixed by choosing a model for \dkpp\ 
decays, the $\tilde{D}$ Dalitz distributions for $B^+$ and $B^-$ decays can be 
fitted simultaneously by the above expressions for $M_{+}$ and $M_{-}$, 
with $r$, $\phi_3$, and $\delta$ as free parameters.
The method is is therefore directly sensitive to the value of $\phi_3$ and 
it does not require additional assumptions on the values of $r$ and $\delta$. 
Moreover, the value of $r$ obtained in the fit can be further used
in other $\phi_3$ measurements. 

To fit the \dtkpp\ Dalitz plot distributions corresponding to 
$B^+$ and $B^-$ decays, we use an unbinned maximum likelihood
technique with the model of neutral $D$ decay determined from 
a flavor-tagged $D$ sample from continuum \dsdpi\ decay. 
The drawback of this approach is that only the absolute value of the 
$D^0$ decay amplitude $f$ is determined directly, but the complex 
form of $f$ can be obtained only 
with certain model assumptions, leading to substantial model uncertainties
in the determination of $\phi_3$. These uncertainties, however, can 
be controlled in future using data from $c\tau$-factories.
The sample of neutral $D$ mesons in a CP eigenstate, which can be produced in the 
decay of $\psi(3770)$ resonance, will provide the information about the 
complex phase of amplitude $f$ which is needed for model-independent 
measurement of $\phi_3$. 

\section{Event selection}

For our measurement we use the 140 fb$^{-1}$ of data collected by 
the Belle detector.\cite{belle} The decays \bdk\ and 
\bdsk, $D^{*0}\to D^0\pi^0$ are selected for the 
determination of $\phi_3$.  
We require $D^0$ to decay to the $K_s\pi^+\pi^-$ final state in all cases.
Also, we select decays of \dsdpi\ produced in the
continuum for the determination of 
the \dkpp\ decay amplitude. 

To determine the $\bar{D^0}$ decay model we use $D^{*\pm}$s 
produced via the $e^+ e^-\to c\bar{c}$ continuum process. 
The flavor of the neutral $D$ meson is tagged by the charge of the slow pion 
$\pi_s$ in the decay \dsdpi.
To select neutral $D$ candidates we require the invariant mass of the 
$K_s\pi^+\pi^-$ system to be within 9 MeV/$c^2$ of the $D^0$ mass, $M_{D^0}$.
To select the events originating from the $D^{*\pm}$ decay 
we make a requirement on the difference 
$\Delta M$ of the invariant masses of
the neutral $D$ and $D^{*\pm}$ candidates: 
$144.6\mbox{ MeV}/c^2<\Delta M<146.4\mbox{ MeV}/c^2$.
To suppress the combinatorial background from $B\bar{B}$ events, 
we require the $D^{*\pm}$ to have momentum 
in the center-of-mass (CM) frame greater than 2.7 GeV/$c$.
The number of events passing selection criteria is 104204.  
The fit of the $\Delta M$ distribution yields the 
background fraction of 3.09$\pm$0.05\%.

Selection of $B$ candidates uses the CM energy difference
$\Delta E = \sum E_i - E_{beam}$ and the beam-constrained $B$ mass
$M_{bc} = \sqrt{E_{beam}^2 - (\sum p_i)^2}$, where $E_{beam}$ is the CM beam 
energy, and $E_i$ and $p_i$ are the CM energies and momenta of the
$B$ candidate decay products. The requirements for signal 
candidates are $5.272$~GeV/$c^2<M_{bc}<5.288$ GeV/$c^2$ 
and $|\Delta E|<0.022$ GeV. 
In addition, we make a  requirement on the invariant mass of the 
neutral $D$ candidate: $|M_{K_s\pi\pi}-M_{D^0}|<11$ MeV/$c^2$. 
To suppress the continuum background, we require $|\cos\theta_{thr}|<0.8$, 
where $\theta_{thr}$ is the angle between the thrust axis of the $B$ candidate 
and the rest of the event. For additional background rejection, we 
use a Fisher discriminant based on ``virtual calorimeter".\cite{fisher} 

The selection efficiency of the \bdk\ process is determined from 
a Monte Carlo (MC) simulation of the detector and amounts to 11\%.
The number of events passing all selection criteria is 146. 
The background fraction extracted from the fit of $\Delta E$
distribution is $25\pm 4$\%.

For the selection of \bdsk\ events in addition to the 
requirements described we require the mass difference 
$\Delta M=M_{K_s\pi^+\pi^-\pi^0}-M_{K_s\pi^+\pi^-}$ of 
neutral $D^{*}$ and $D$ candidates to satisfy 
$140\mbox{ MeV}/c^2<\Delta M<145\mbox{ MeV}/c^2$.
The selection efficiency is 6.2\%, and the number of events 
satisfying selection criteria is 39. 
The background fraction is $12\pm 4$\%.

\section{Determination of \boldmath{\dkpp} decay model}

The amplitude $f$ of the \dkpp decay is represented
by a coherent sum of two-body decay matrix elements, each having its
own amplitude and phase, plus one non-resonant 
decay amplitude. The total phase and amplitude 
are arbitrary. To be consistent with the CLEO analysis,\cite{dkpp_cleo} 
we have chosen the $K_s\rho$ 
mode to have unit amplitude and zero relative phase. 
The description of the matrix elements follows Ref.~\cite{cleo_model} 

For our $D^0$ model fit we use a set of 15 two-body amplitudes. 
These include four Cabibbo-allowed amplitudes: $K^*(892)^+\pi^-$, $K_0^*(1430)^+\pi^-$, 
$K_2^*(1430)^+\pi^-$ and $K^*(1680)^+\pi^-$, their
doubly Cabibbo-suppressed partners, and seven 
channels with $K_s$ and a $\pi\pi$ resonance: 
$K_s\rho$, $K_s\omega$, $K_sf_0(980)$, $K_sf_2(1270)$, 
$K_sf_0(1370)$, $K_s\sigma_1$ and $K_s\sigma_2$. The masses and Breit-Wigner 
widths of scalars $\sigma_1$ and $\sigma_2$ are left unconstrained, while 
the parameters of other resonances are taken to be the same as in the CLEO 
analysis.\cite{dkpp_cleo} The parameters of the $\sigma$ resonances obtained 
in the fit are as follows: $M_{\sigma_1}=539\pm 9$ MeV/$c^2$, 
$\Gamma_{\sigma_1}=453\pm 16$ MeV/$c^2$, $M_{\sigma_2}=1048$ MeV/$c^2$, 
$\Gamma_{\sigma_2}=109\pm 11$ MeV/$c^2$. The resonance $\sigma_2$ was 
introduced to describe a structure in the Dalitz distribution possibly due to 
the decay $f_0(980)\to\eta\eta$ with the rescattering of 
$\eta\eta$ to $\pi^+\pi^-$. 

We use the unbinned maximum likelihood technique to fit the Dalitz plot 
distribution. The fit function for the Dalitz plot density 
is represented by a sum of the squared absolute value of the 
decay amplitude $|f(m^2_+, m^2_-)|^2$ and the background distribution, 
convoluted with a momentum resolution function and multiplied by the 
function describing efficiency. The free parameters of the minimization 
are the amplitudes and phases of the resonances, and the amplitude 
and phase of the non-resonant component. 
The background density for \dkpp\ events is obtained from 
$\Delta M$ sidebands. 
The shape of the efficiency over the Dalitz plot is extracted from a
MC simulation. The fit result is presented in Table~\ref{dkpp_table}. 

\begin{table}
\caption{Fit results for \dkpp\ decay. Errors are statistical only.}
\label{dkpp_table}
\centering
\small
\begin{tabular}{|l|c|c|c|c|c|} \hline
Intermediate state           & Amplitude & Phase ($^{\circ}$) \\ \hline

$K^*(892)^+\pi^-$            & $1.656\pm 0.012$
                             & $137.6\pm 0.6$ 
                             \\ 

$K^*(892)^-\pi^+$            & $(14.9\pm 0.7)\times 10^{-2}$
                             & $325.2\pm 2.2$
                             \\

$K_0^*(1430)^+\pi^-$         & $1.96\pm 0.04$
                             & $357.3\pm 1.5$
                             \\

$K_0^*(1430)^-\pi^+$         & $0.30\pm 0.05$
                             & $128\pm 8$
                             \\

$K_2^*(1430)^+\pi^-$         & $1.32\pm 0.03$
                             & $313.5\pm 1.8$
                             \\

$K_2^*(1430)^-\pi^+$         & $0.21\pm 0.03$
                             & $281\pm 9$
                             \\

$K^*(1680)^+\pi^-$           & $2.56\pm 0.22$
                             & $70\pm 6$
                             \\

$K^*(1680)^-\pi^+$           & $1.02\pm 0.2$
                             & $103\pm 11$
                             \\

$K_s\rho^0$                  & $1.0$ (fixed)                                 
                             & 0 (fixed)   
                             \\

$K_s\omega$                  & $(33.0\pm 1.3)\times 10^{-3}$
                             & $114.3\pm 2.3$
                             \\

$K_s f_0(980)$               & $0.405\pm 0.008$
                             & $212.9\pm 2.3$
                             \\

$K_s f_0(1370)$              & $0.82\pm 0.10$ 
                             & $308\pm 8$
                             \\

$K_s f_2(1270)$              & $1.35\pm 0.06$
                             & $352\pm 3$
                             \\

$K_s \sigma_1$               & $1.66\pm 0.11$
                             & $218\pm 4$
                             \\

$K_s \sigma_2$               & $0.31\pm 0.05$
                             & $236\pm 11$
                             \\

non-resonant                 & $6.1\pm 0.3$ 
                             & $146\pm 3$ 
                             \\ 
\hline
\end{tabular}
\end{table}

\section{Dalitz plot analysis of \boldmath{\bddsk} decay}

The Dalitz plots of $\tilde{D}$ decaying to $K_s\pi^+\pi^-$, 
which contain information about CP violation in $B$
decays, are fitted by minimizing the combined logarithmic 
likelihood function for $B^-$ and $B^+$ data sets. 
The corresponding Dalitz plot densities are based on
decay amplitudes $M_{\pm}$ described by Eq.~\ref{intdist} ($B^+$ data)
and \ref{intdist_m} ($B^-$ data). 
The $\bar{D^0}$ decay model $f$ is fixed, and the free parameters of the
fit are the amplitude ratio $r$ and phases $\phi_3$ and $\delta$. 

Since the background rate is significant, it is essential to include 
the background density into the fit model. 
The largest background contribution comes from 
continuum $e^+e^-\to q\bar{q}$ ($q=u, d, s, c$) events. 
It includes background with purely 
combinatorial tracks, and continuum $D^0$ mesons combined with a
random kaon. This type of background is analyzed using an event sample 
in which $\cos\theta_{thr}$
and Fisher discriminant requirements are applied in order to 
select continuum events. The continuum background fraction is 
$22.1\pm 3.9$\% for \bdk\ mode and $9.0\pm 3.6$\% for \bdsk\ mode. 
The background due to $B\bar{B}$ combinatorics was investigated 
using the MC simulation. 
The fraction of $B\bar{B}$ background is $3.6\pm 0.3$\% for \bdk\ mode and 
$3.1\pm 0.4$\% for \bdsk\ mode. 

To test the consistency of the fit, the same procedure was 
applied to the \bddstpi\ and $\bar{B^0}(B^0)\to D^{*\pm}\pi^{\mp}$ 
control samples as to the \bddstk\ signal. 
For decays of flavor eigenstate of $D$ meson, our fit 
should return $r$ values consistent with zero.
In the case of \bddstpi\, a small amplitude ratio is expected 
($r\sim |V_{ub} V^*_{cd}|/|V_{cb}V^*_{ud}|\sim 0.02$). Deviations 
from these values can appear if the Dalitz plot shape is not well 
described by the fit model. 
For the control sample fits, we consider $B^+$ and $B^-$ data separately, 
to check for the absence of CP violation. 
In the fit of \bdtpi\ data, we obtain $r\sim 0.06$ which is more than 
two standard deviations apart from zero, however, no CP asymmetry is found. 
This bias is considered as a systematic effect. 
Other control samples, \bdstpi\ with $D^{*0}$ decaying to 
$D^0\pi^0$, and $\bar{B^0}(B^0)\to D^{*\pm}\pi^{\mp}$ with 
$D^{*\pm}\to D^0\pi^\pm$, do not show any significant deviation from $r=0$. 

The combined unbinned maximum likelihood fit of 
$B^+$ and $B^-$ samples with free parameters
$r$, $\phi_3$ and $\delta$ yields the following values: 
$r=0.31\pm 0.11$, $\phi_3=86^{\circ}\pm 17^{\circ}$, 
$\delta=168^{\circ}\pm 17^{\circ}$ for the \bdtk\ sample and 
$r=0.34\pm 0.14$, $\phi_3=51^{\circ}\pm 25^{\circ}$, 
$\delta=302^{\circ}\pm 25^{\circ}$ for the \bdstk\ sample.
The method has a two-fold ambiguity ($\phi_3+\pi$, $\delta+\pi$), 
since this transformation does not change the total phases, 
which are actually measured. Here we choose the solution with 
$0<\phi_3<\pi$. 

The errors presented above are the estimation obtained from the likelihood fit.
For a more reliable estimation of the statistical errors 
we use a Bayesian approach with the probability density function (PDF)
of the fitted parameters obtained from a large number of MC 
pseudo-experiments. 
The MC procedure consists of a generation of two Dalitz plot
samples of $D^0$-$\bar{D}^0$ mixture (with total opposite flavor phases
$\delta-\phi_3$ and $\delta+\phi_3$) with the efficiency, 
momentum resolution and 
background taken into account as in the fit of experimental data, 
with a subsequent fitting of this samples to extract the values of 
$r$, $\phi_3$ and $\delta$. The number of events in the samples is taken
nearly equal to the number of events in the experimental data.
This procedure was repeated in several hundred trials for different 
values of input $r$ parameter. The PDF is then parameterized to obtain the 
function describing
the probability density of the reconstructed parameters for any
set of true parameters. After the fitted parameters PDF is obtained, 
we assume a flat prior PDF and calculate the PDF for the true 
parameters. 
This procedure not only allows to obtain more reliable
estimation of errors, but also corrects the systematic bias of the 
value of $r$ introduced by the fit procedure due to its positive
definiteness. 

\begin{figure}
  \vspace{-2\baselineskip}
  \centering\epsfig{figure=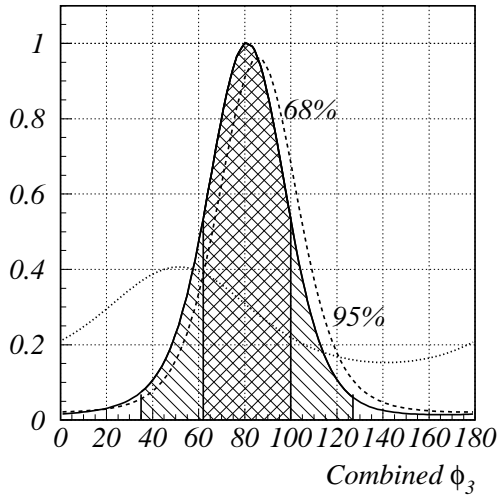,width=0.47\textwidth}
  \caption{$\phi_3$ probability density function for \bdk\ (dashed line), 
           \bdsk\ (dotted line) and combined measurement (solid line).}
  \label{fcomb}
\end{figure}

The result of the fit with errors (68\% confidence level) obtained from the ``toy" MC are 
as follows: $r=0.28^{+0.09}_{-0.11}$, $\phi_3=86\pm 20^{\circ}$, 
$\delta=168\pm 20^{\circ}$ for \bdk\ and $r<0.25
$, 
$\phi_3=51\pm 47^{\circ}$, $\delta=302\pm 47^{\circ}$ for \bdsk. The 
95\% confidence intervals are $0.07<r<0.45$, $37^{\circ}<\phi_3<135^{\circ}$, 
$119^{\circ}<\delta<217^{\circ}$ for \bdk\ and 
$r<0.44$, $-31^{\circ}<\phi_3<133^{\circ}$, 
$220^{\circ}<\delta<384^{\circ}$ for \bdsk\ . The significance of CP violation 
is 94\% for \bdk\ and 38\% for \bdsk. Though the 68\% CL error is quite small, 
the significance is comparatively low due to longer tails of the PDF 
compared to the gaussian distribution. 

We produce the combined measurement of $\phi_3$ by multiplying the 
$\phi_3$ PDFs for \bdk\ and \bdsk\ modes. The resulting PDF, as well as
the PDFs for individual measurements are shown in Fig.~\ref{fcomb}.
The result of the combined measurement is $\phi_3=81\pm 19^{\circ}$, 
the 95\% confidence interval is $35^{\circ}<\phi_3<127^{\circ}$. 

The model used for the \dkpp\ decay is one of the main sources of 
systematic errors for our analysis. The model is a result of 
the experimental Dalitz plot fit, but since the plot density is
proportional to the squared absolute value of the decay amplitude, 
the phase $\phi(m^2_+, m^2_-)$ of the complex amplitude is not 
directly measured in the experiment. The phase dependence is
therefore the result of model assumptions and 
its uncertainty may affect the $\tilde{D}$ Dalitz plot fit. 
To estimate 
the model uncertainties, a MC simulation is performed. 
Event samples are generated according to the amplitude 
described by Eq.~\ref{intdist} 
with the resonance parameters extracted from 
our fit of continuum $D^0$ data, but to fit this distribution
different models for $f(m_+, m_-)$ are used. We scan the phases $\phi_3$
and $\delta$ in their physical regions and take the maximum 
deviation of the fit parameter as a model uncertainty estimation. 
Since the 
Breit-Wigner amplitude, on which the $\bar{D^0}$ model is based, 
can describe well only the narrow resonances, 
our estimation of the model uncertainty is based on the fit model 
containing only the narrow resonances,
with the wide ones approximated by the constant complex term.
The estimated value of the systematic uncertainty on $\phi_3$ is $11^{\circ}$.

Other sources of systematic errors include the uncertainties of the
knowledge of the detector response, background estimation 
and possible fit biases.
The component related to the background 
shape parameterization is estimated by extracting the background shape 
from the $M_D$ sidebands and by using a flat background distribution.
To estimate the contributions of efficiency shape evaluation and 
momentum resolution to the systematic error, we 
repeat the fit using a flat efficiency and a fit model which
does not take the resolution into account, respectively. 

The non-zero amplitude ratio observed in the \bdpi\ control sample
can be either due to the statistical fluctuation or 
may indicate some systematic effect such as background structure or a deficiency 
of the $\bar{D^0}$ decay model. Since the source of this bias is indeterminate, 
we conservatively treat it as an additional systematic effect. 
The corresponding bias of the weak and strong phases is $11^{\circ}$. 
This contribution dominates in the systematic error, which equals to 
$13^{\circ}$ for \bdk\ and $11^{\circ}$ for \bdsk\ mode. 

\section{Conclusion}

We have studied a new method to measure the unitarity triangle angle $\phi_3$ 
using Dalitz plot analysis of the three-body $D^0$ decay in the process 
\bddsk. The first measurement of $\phi_3$ using this technique
was performed based on 140 fb$^{-1}$ statistics collected by 
the Belle detector. 
From the combined fit of \bdk\ and \bdsk\ modes, we obtain the 
value of $\phi_3=81^{\circ}\pm 19^{\circ}\pm 13^{\circ}\pm 11^{\circ}$.
The first error is statistical, the second is experimental systematics and
the third is model uncertainty. 
The 95\% confidence interval is $35^{\circ}<\phi_3<127^{\circ}$
The statistical significance of the CP violation is 94\% for \bdk\ 
mode and 38\% for \bdsk\ mode. 

The method has a number of advantages over the other
ways to measure $\phi_3$.\cite{bdk_gamma} 
It is directly sensitive to the value of $\phi_3$ and has only the two-fold 
discrete ambiguity ($\phi_3+\pi$, $\delta+\pi$). 
It does not involve branching 
fraction measurements and, therefore, the influence of the detector 
systematics is lower. Also, the statistical power of this technique is 
higher in the presence of the background since the interference 
term is measured. 

\section*{Acknowledgments}

We are grateful to V.~Chernyak and M.~Gronau for fruitful 
discussions. 
We wish to thank the KEKB accelerator group for the excellent
operation of the KEKB accelerator.
We acknowledge support from the Ministry of Education,
Culture, Sports, Science, and Technology of Japan
and the Japan Society for the Promotion of Science;
the Australian Research Council
and the Australian Department of Education, Science and Training;
the National Science Foundation of China under contract No.~10175071;
the Department of Science and Technology of India;
the BK21 program of the Ministry of Education of Korea
and the CHEP SRC program of the Korea Science and Engineering Foundation;
the Polish State Committee for Scientific Research
under contract No.~2P03B 01324;
the Ministry of Science and Technology of the Russian Federation;
the Ministry of Education, Science and Sport of the Republic of Slovenia;
the National Science Council and the Ministry of Education of Taiwan;
and the U.S.\ Department of Energy.

\end{document}